\documentclass{elsart}
\usepackage{graphicx}
\usepackage{amsmath,amssymb}
\begin{document}

\setlength\paperwidth{8.5in}
\setlength\paperheight{11in}
\setlength{\pdfpagewidth}{\paperwidth}
\setlength{\pdfpageheight}{\paperheight}
\interfootnotelinepenalty=100000

\begin{frontmatter}

\title{Time-Dependent Point Source Search Methods in High Energy Neutrino Astronomy}

\author[label1]{Jim Braun}\footnote{Corresponding author: jim.braun@icecube.wisc.edu}, 
\author[label1]{Mike Baker},
\author[label1]{Jon Dumm}, 
\author[label1]{Chad Finley}, 
\author[label1]{Albrecht Karle},
\author[label1]{Teresa Montaruli}
\address[label1]{University of Wisconsin, Chamberlin Hall, Madison, Wisconsin 53706, USA}

\begin{abstract}
We present maximum-likelihood search methods for time-dependent fluxes from point sources, such as flares or periodic emissions.
We describe a method for the case when the time dependence of the flux
can be assumed {\it a priori} from other observations, and we additionally describe a method to search for bursts with an unknown time dependence.
In the context of high energy neutrino astronomy, we simulate one year of data from a cubic-kilometer scale neutrino
detector and characterize these methods and equivalent binned methods with respect to the duration of neutrino emission.
Compared to standard time-integrated searches, we find that up to an order of magnitude fewer events are needed to discover bursts with short durations,
even when the burst time and duration are not known {\it a priori}.

\end{abstract}
\end{frontmatter}

\section{Introduction}

High energy gamma ray astronomy experiments, such as the imaging atmospheric Cherenkov 
telescopes (IACTs) H.E.S.S., MAGIC, VERITAS and CANGAROO, and extensive air shower arrays such as Milagro, Argo, and Tibet Array, have revealed a large number of
sources with emissions extending to above a few tens of TeV (for a collection of results see \cite{tevcat}). The Fermi Gamma-ray Space Telescope, launched in June 2008,
has already produced first catalogues for sources emitting $>100$~MeV photons \cite{fermibsl}. TeV gamma ray sources include galactic supernova remnants, pulsars, and
microquasars, as well as extragalactic gamma ray bursts (GRBs) and active galactic nuclei (AGN).  Gamma ray fluxes produced by many of these sources are time dependent.
High energy photon observations of AGN reveal flaring activities on timescales of hours to several weeks, with intensities often several times larger than the typical flux
of the source in its quiescent state. These flares are yet unpredictable. GRBs have burst timescales ranging from milliseconds to a few minutes \cite{kouv} that are still not
fully explained.  On the other hand, some of the time dependent emissions are more predictable. For instance, binary systems are naturally periodic. TeV photon fluxes from the
microquasars LS I +61 303 and LS 5039, for example, are modulated by the orbital phase of the system \cite{magiclsi, hessls50}. 

No high energy extraterrestrial neutrino sources have been discovered, but potential sources include high energy gamma ray sources.  Neutrino fluxes can be similarly time
dependent if neutrinos and gamma rays originate from meson decays produced in hadronic processes (interactions of protons or nuclei with matter or environmental photons near sources).
During AGN flares, it is possible that an enhanced emission on top of that produced by electromagnetic mechanisms, such as synchrotron radiation and inverse Compton, is due to an
increase in proton acceleration efficiency in jets \cite{reimer}.  The periodicity of microquasar photon emissions suggests that any neutrino emission may be similarly periodic.

The central statistical challenge in high energy astronomy is the identification of event excesses due to sources amongst random background fluctuations.
Features that differentiate the signal emission from the background are used in this process, namely directional and energy information.
Another distinguishing feature is time.  For example, if a burst is observed in a photon experiment, one can test for a coincident neutrino emission.
A much better background rejection, and thus a better sensitivity, is achieved by selecting only events near the time of the flare or burst.  
Furthermore, one may wish to search for time dependent sources without a corresponding trigger from another experiment by testing if a cluster of events in time is
incompatible with the random time distribution of the background.
We call these searches untriggered, since they do not use information from other experiments.  Such untriggered searches (e.g. \cite{konstancja, hughey, vasileiou})
use time-variable or rolling time bins to identify excesses, and generally outperform time-independent searches if the signal events are indeed clustered in time.
In Ref.~\cite{method}, we demonstrated that binned methods are not as effective as unbinned methods based on maximum likelihood techniques, which describe the signal and background using probability density functions (PDFs), and we applied this method to a simulated search for high energy neutrino point sources in a Cherenkov neutrino telescope. Such methods have been applied in astroparticle point source searches \cite{aguilar, negrb, thrane, alexandreas, amanda, ic22}.  

Here, we extend this method to search for point sources with time-dependent fluxes, incorporating a time-dependent
term into the signal and background PDFs.  We first review binned methods in Sec.~\ref{sec:bin}, and in Sec.~\ref{sec:method} we describe an unbinned method
to search for bursts, flares, and other cases when the nature of the time dependence can be assumed from observations of other experiments.  We then
describe a method for an untriggered search, when the burst time and duration are not known.  In sections \ref{sec:results} and \ref{sec:results2}, we
apply the methods to a simulated neutrino search and characterize the discovery potential of these methods as a function of the burst duration.

\section{Binned Methods}
\label{sec:bin}

The data in high energy astronomy consist of a set of events distributed throughout a region of the sky, and can be modeled by two hypotheses:  Either the data consist solely of background events, i.e. the null hypothesis, or the data additionally contains a signal from an astrophysical
source.  The signal and background event distributions are governed by PDFs, describing the event angular distribution, energy spectrum, time distribution,
etc.  Any difference between the signal and background PDFs provides an opportunity to differentiate events produced by an astrophysical source from the background.
Binned and unbinned methods use these PDFs in different ways.  For example, events from an astrophysical point source cluster around the location of
the source with an angular deviation given by the detector point spread function, which often is approximately a 2-D Gaussian.  When using binned methods to evaluate whether a source is
present at a given location $\vec{x}_s$ in an event sky map, $N$ events can be selected within an angular bin comparable in size to the detector angular
resolution and centered on $\vec{x}_s$, and $N$ is then compared to the expected background using Poisson or binomial statistics.  If the probability of obtaining
$N$ or more events from background alone is less than a given confidence level threshold, e.g. 5$\sigma$ ($p$=5.73$\times$10$^{-7}$), we reject the null
hypothesis and decide that a source is present in the data.  The mean number of signal events necessary to reject the null hypothesis in a given fraction of
trials (e.g. 50\%) at the given CL is the discovery potential, a figure of merit for the search.

Additional information may be present in other variables characterizing the data.  Particularly, if the signal events are clustered on a small timescale of
approximately $\Delta$T relative to the duration of the experiment T, applying a time selection in addition to the angular selection reduces
the background by a factor $\sim$$\Delta$T/T and can significantly improve the discovery potential.  A gain is similarly possible with an energy-based
selection if the energy spectra of potential sources are significantly different from the background.

Binned methods are easy to implement and computationally fast.  Criticisms of binned methods and other methods incorporating an event selection are
generally the following:
\begin{itemize}
\item {\bf The information reduction problem}.  All of the event information is reduced to a binary classification; either the event passes the
selection and is counted, or it does not.  A fraction of potential signal events is always lost.  Additionally, information contained within the event
distribution is lost that alternatively could indicate the relative agreement of each event with signal or background.  For example, events at the edge
of a search bin are not as indicative of a point source as events near the center, but are counted the same.  Finally, classes of signal events may have
substantially different angular resolution; e.g. in many cases in astroparticle physics, high energy events are reconstructed more accurately.
By applying the same angular bin size for all events, binned methods fail to dynamically incorporate such information.
\smallskip
\item {\bf The optimization problem}.  The event selection, including the angular bin radius, must be optimized given a specific point source signal
hypothesis.  For instance, a binned analysis optimized to discover fluxes with hard energy spectra may be far from optimal for sources with softer energy spectra.
The selection often needs to be reoptimized to test a different hypothesis, which can be problematic, particularly when the analysis is performed
adhering to blindness principles.  Additionally, the selection
which optimizes sensitivity \cite{mrf} (i.e. set the best limits) generally does not maximize discovery potential, and therefore usually one or the other is sacrificed.
\end{itemize}

\section{Unbinned Methods}
\label{sec:method}

Unbinned methods model the data as a two component mixture of signal
and background and fit the data to determine the relative contribution of each component.  The general procedure is as follows.  Events are selected from
a region surrounding the source, considerably larger than the detector angular resolution.  Given
$N$ events in the data set, the probability density of the i$^{th}$ event is
\begin{equation}
\frac{n_s}{N}\mathcal{S}_i + (1 - \frac{n_s}{N})\mathcal{B}_i,
\end{equation}
where $\mathcal{S}_i$ and $\mathcal{B}_i$ are the signal and background PDFs, respectively. The parameter $n_s$ is the unknown contribution
of signal events.
The likelihood of the data given $n_s$ is the product of the event probability densities:\footnote{The likelihood has alternatively
been derived with a term describing the Poisson probability of obtaining $N$ events \cite{negrb, thrane}.  This term is not necessary when the background
is determined from off-source data and sufficient off-source events are included in $N$.}
\begin{equation}
\mathcal{L}(n_s) = \prod_{i=1}^{N} \Big[\frac{n_s}{N}\mathcal{S}_i + (1 - \frac{n_s}{N})\mathcal{B}_i\Big].
\label{eq:lh}
\end{equation}
The likelihood is then maximized with respect to $n_s$ and other free parameters, giving the best fit value $\hat{n}_s$.  The maximization can
be done numerically with e.g. MINUIT \cite{minuit} or equivalently with the expectation-maximization technique \cite{aguilar}.  The null hypothesis
is given by $n_s=0$ and is nested within the allowed parameter space.  The standard likelihood ratio test statistic is
\begin{equation}
D = -2 \log \lambda = -2 \log\Big[\frac{\mathcal{L}(n_s=0)}{\mathcal{L}(\hat{n}_s)}\Big] \times \text{sign}(\hat{n}_s),
\label{eq:ts}
\end{equation}
where the factor $\text{sign}(\hat{n}_s)$ differentiates between negative and positive excesses.  The obtained value of $D$ is then compared
to the distribution of $D$ given the null hypothesis.  Large values of $D$ reject the null hypothesis with a confidence level equal to
the fraction of the distribution above the obtained $D$, and discovery potential is calculated as described in section \ref{sec:bin}.

\subsection{A Point Source with an Assumed Time Dependence}
\label{sec:mtd1}

Suppose we wish to test for the existence of a neutrino point source at a given location $\vec{x}_s$.  We assume the neutrino emission from
the source should follow a time distribution with a known functional form.  The unbinned signal PDF is the product of independent space angle,
energy, and time probability terms:
\begin{equation}
\mathcal{S}_{i} = \mathcal{N}(r_i) \times \mathcal{E}(\text{E}_i) \times \mathcal{T}(\text{T}_i).
\end{equation}
For an event with reconstructed direction $\vec{x}_i$, we model the probability of originating from the source, with a space angle difference
$r_i = |\vec{x}_i - \vec{x}_s|$, as a 2-D Gaussian:
\begin{equation}
\mathcal{N}(r_i) = \frac{1}{2\pi \sigma_i^2}e^{-{\frac{r_i^2}{2\sigma_i^2}}},
\end{equation}
where $\sigma_i$ is the angular resolution of the 2-D Gaussian (i.e. 39.3\% of $r_i$ are equal to or less than $\sigma_i$).  In many
applications, $\sigma_i$ can be determined for each event individually \cite{till}, or alternatively, $\sigma_i$ can be determined from
Monte Carlo simulations.  Since angular resolution is generally energy dependent, the ability to use individual $\sigma_i$ ensures the independence of
the spatial and energy terms in the signal PDF.

The energy PDF $\mathcal{E}(\text{E}_i)$ describes the probability of obtaining a reconstructed energy E$_i$ for an event produced by a source of a given energy spectrum.
The typical expectation of an E$^{-2}$ power-law energy spectrum for astrophysical neutrino sources (characteristic of Fermi acceleration mechanisms)
results in an excess of higher energy signal events compared with the background of atmospheric neutrinos ($\sim$E$^{-3.6}$  spectrum above 100 GeV).   This information
makes it easier to identify a signal for harder spectra, and therefore fewer events are needed for a significant detection.  We assume the source energy spectrum
follows a power law $\frac{dN}{d\text{E}}$$\sim$E$^{-\gamma}$ with spectral index $\gamma$.
Since we do not know the exact spectral index of the source, we allow $\gamma$ to be a free parameter in our model. As in \cite{method}, we generate normalized probability
tables of $\mathcal{E}(\text{E}_i|\gamma)$, the probability of obtaining a reconstructed energy E$_i$ for an event produced by a source with spectral index $\gamma$,
for spectral indices 1.0 $<$ $\gamma$ $<$ 4.0 using Monte Carlo.  We then fit $\gamma$ along with $n_s$ in the likelihood maximization.  We have found that the use of
the energy term improves the discovery potential significantly in time-independent neutrino point source searches, and, if enough signal
events are present, the best fit spectral index $\hat{\gamma}$ provides a reasonable measurement of the source spectral index \cite{method}.

The time PDF $\mathcal{T}(\text{T}_i)$ describes the time distribution of events observed from the source.  In this case, we assume the time distribution
from photon observations.  For example, we could search for the high energy neutrino counterpart to an orphan TeV photon flare\footnote{Orphan means that no coincident
X-ray emission is observed.} of an AGN by using the lightcurve measured by an IACT during the flare as a PDF,
or we can look for high energy neutrinos in coincidence with MeV~--~GeV photon observations of a GRB.
Alternatively, we can use long-term AGN lightcurves to investigate if similar variations of 
neutrino and photon fluxes are observed, for example using Fermi-LAT daily monitoring.
 
We consider, as an example, flares with Gaussian time dependence:
\begin{equation}
\mathcal{T}(\text{T}_i) = \frac{1}{\sqrt{2\pi} \sigma_{\text{T}}}e^{-{\frac{(\text{T}_i-\text{T}_{\circ})^2}{2\sigma_{\text{T}}^2}}} \times \mathcal{H}(\text{T}_i),
\end{equation}
where $\text{T}_{\circ}$ and $\sigma_{\text{T}}$ are the Gaussian mean time and Gaussian width of the burst, respectively, which are assumed to be known.
We add the term $\mathcal{H}(\text{T}_i)$, which describes the detector efficiency at the source location as a function of time.  Particularly, detector efficiency
is often dependent on zenith and azimuth angles and is zero when the detector is off or $\vec{x}_s$ is outside the detector field of view.  The time PDF
$\mathcal{T}(\text{T}_i)$ must be normalized to unity while including this efficiency term.  For this work, we consider uniform efficiency, i.e. $\mathcal{H}(\text{T}_i)$=1.

Including all terms, the signal PDF is:
\begin{equation}
\mathcal{S}_i = \frac{1}{2\pi \sigma_i^2}e^{-{\frac{r_i^2}{2\sigma_i^2}}}\times \mathcal{E}(\text{E}_i|\gamma) \times \frac{1}{\sqrt{2\pi} \sigma_{\text{T}}}e^{-{\frac{(\text{T}_i-\text{T}_{\circ})^2}{2\sigma_{\text{T}}^2}}}.
\label{eq:sigpdf}
\end{equation}
The background PDF $\mathcal{B}_i$ contains the same three terms, describing the angular, energy, and time distributions of background events:
\begin{equation}
\mathcal{B}_i = \frac{1}{\Omega \text{T}_L} \mathcal{E}(\text{E}_i|Atm_{\nu}) \times \mathcal{H}(\text{T}_i),
\label{eq:bgpdf}
\end{equation}
where $\mathcal{E}(\text{E}_i|Atm_{\nu})$ is the probability of obtaining $\text{E}_i$ from atmospheric neutrinos, T$_L$ is the livetime of the data set, and $\Omega$ is the solid
angle of a declination band centered on the declination of the source and containing $N$ total events (in this work, we have used bands of $\pm$6$^{\circ}$).  We again
assume uniform efficiency over time, with $\mathcal{H}(\text{T}_i)$=1.
We maximize the likelihood of Eq. \ref{eq:lh} with respect to both the number of signal events $n_s$ and the spectral index
$\gamma$, yielding best fit values $\hat{n}_s$ and $\hat{\gamma}$ and test statistic:
\begin{equation}
D = -2 \log\Big[\frac{\mathcal{L}(n_s=0)}{\mathcal{L}(\hat{n}_s, \hat{\gamma})}\Big] \times \text{sign}(\hat{n}_s).
\label{eq:knts}
\end{equation}

\subsection{A Point Source with Unknown Time Dependence}
\label{sec:mtd2}

A transient neutrino signal may not be triggered by photon observations, for example if the source is opaque to photons or if the neutrino emission is offset with respect to the
photon emission. To identify such untriggered transient signals, it is necessary to search the data for excesses with respect to time.  Binned methods designed to detect
untriggered transient signals are well known.  For example, if the duration of the burst is assumed, a sliding time window can be used \cite{hughey}.
More generally, the burst duration is not known and must be determined along with the time of occurrence.  Plausible burst
durations may span several orders of magnitude.  One approach to this problem is to use multiple sliding time windows with different
durations \cite{vasileiou}.  In another approach \cite{konstancja}, all possible groups of consecutive events within the angular search
bin are considered, and the most unlikely group (with respect to the background time distribution) is chosen as a burst candidate.  All searches for bursts with an unknown
time have a large trial factor on the significance of the final result.

The functional form of the time dependent signal is additionally not known.
Nonetheless, low statistics signals producing a small number of recorded events contain little information on the details of these functions and can be
fit equally well by a range of functional forms.  Because a Gaussian provides a reasonable and convenient fit to a generic transient signal, we therefore search the data for a
point source with a Gaussian time PDF, and the signal and background PDFs are identical to Eqs. \ref{eq:sigpdf} and \ref{eq:bgpdf}, respectively.
In this case, the Gaussian mean time $\text{T}_{\circ}$ and width $\sigma_{\text{T}}$ are unknown, along with the number of signal events $n_s$ and spectral index $\gamma$.

When fitting also for the unknown burst time $\text{T}_{\circ}$ and width $\sigma_{\text{T}}$, we need to address a subtlety that emerges in the approach of maximizing
the likelihood with respect to the unknown parameters as was done in Sec. \ref{sec:mtd1}.  The effective trial factor, describing the number of independent ways to choose the burst
time within the time window, is dependent on the burst duration, with a larger trial factor for shorter durations.  The simple likelihood maximization does not account for
this effect and favors bursts with shorter durations.

We correct this behavior by marginalizing with respect to the burst time in a Bayesian manner, using a uniform prior.  For a search window bounded in time by $\text{T}_{min}$ and
$\text{T}_{max}$, the likelihood is
\begin{equation}
\mathcal{L}(n_s, \gamma, \sigma_{\text{T}}) = \int_{\text{T}_{min}}^{\text{T}_{max}} \prod_{i=1}^{N} \Big[\frac{n_s}{N}\mathcal{S}_i + (1 - \frac{n_s}{N})\mathcal{B}_i\Big] P(\text{T}_{\circ}) d\text{T}_{\circ},
\label{eq:mglh}
\end{equation}
where $P(\text{T}_{\circ})$ = $\frac{1}{\text{T}_{max} - \text{T}_{min}}$ is a constant prior for the time window, and the PDFs
$\mathcal{S}_i$ and $\mathcal{B}_i$ are given by Eqs. \ref{eq:sigpdf} and \ref{eq:bgpdf}, respectively.  If the data contains a significant burst, only times within
$\sim$$\hat{\sigma}_{\text{T}}$ of the burst contribute to the integral, shown in Fig. \ref{Fig:TS}.  The integrand has Gaussian dependence with
respect to $\text{T}_{\circ}$, and the maximum is the maximized likelihood $\mathcal{L}(\hat{n}_s, \hat{\gamma}, \hat{\sigma}_{\text{T}}, \hat{\text{T}}_{\circ})$.
The marginal likelihood in Eq. \ref{eq:mglh} can therefore be approximated by
\begin{equation}
\mathcal{L}(\hat{n}_s, \hat{\gamma}, \hat{\sigma}_{\text{T}}) \sim \frac{\sqrt{2 \pi}\hat{\sigma}_{\text{T}}}{\text{T}_{max} - \text{T}_{min}} \mathcal{L}(\hat{n}_s, \hat{\gamma}, \hat{\sigma}_{\text{T}}, \hat{\text{T}}_{\circ}),
\end{equation}
up to a small factor, which we ignore since it will cancel in the significance calculation described in Sec. \ref{sec:resut}.
Our test statistic is (ignoring also the factor of $\sqrt{2 \pi}$):
\begin{equation}
D = -2 \log\Big[\frac{\text{T}_{max} - \text{T}_{min}}{\hat{\sigma}_{\text{T}}} \times \frac{\mathcal{L}(n_s=0)}{\mathcal{L}(\hat{n}_s, \hat{\gamma}, \hat{\sigma}_{\text{T}}, \hat{\text{T}}_{\circ})}\Big] \times \text{sign}(\hat{n}_s).
\label{eq:uknts}
\end{equation}
The approximation allows us to use the simple likelihood maximization, which is computationally much faster than integration.
We include the factor $\text{sign}(\hat{n}_s)$, although $\hat{n}_s$ is rarely negative since a potential burst is nearly always identified.

Finally, the methods of Sec. \ref{sec:mtd1} and \ref{sec:mtd2} can be similarly used to search for periodic emissions from sources with well-known periodicity by searching
with respect to the orbital phase of the system rather than time.  For example, the microquasars LS I +61 303 and LS 5039 are binary systems where one member is a compact object
with an accretion disk and relativistic jets.  A component of the TeV photon emissions from these sources is modulated according to the orbital phase of the system
\cite{magiclsi, hessls50}.  The phase of a corresponding neutrino component may differ from the photon phase, however, since photons may be absorbed when the companion star
obscures the accelerating compact object \cite{Aharonian:2005cx,Torres:2006ub}.  A signal PDF for such a neutrino search would be similar to Eq. \ref{eq:sigpdf},
substituting the best fit orbital phase and phase width for $\text{T}_{\circ}$ and $\sigma_{\text{T}}$, respectively.

\section{Application to a Simulated Neutrino Point Source Search}
\label{sec:results}

We apply the methods described in the previous section to a simulated search for high energy neutrino bursts.  Our simulation is a reasonable approximation to data
expected from a cubic kilometer scale neutrino observatory such as IceCube \cite{icecube} or a future experiment in the Mediterranean \cite{km3net, nemo}, and is identical
to that described in \cite{method}.  We simulate one year of livetime for a detector at the South Pole, with a cubic kilometer of instrumented ice volume. The detector is composed of 81 strings arranged in a square grid with 125 m spacing of nearest neighbors.  Each string contains 60 optical modules
vertically spaced by 17 m.  Neutrino fluxes are simulated according to \cite{nsim}, using the CTEQ6 structure functions \cite{cteq6} and Preliminary Reference Earth Model \cite{prem}.
Muons are propagated to the detector using MUM \cite{mum}, with energy losses described by dE/dx~=~a~+~bE, with a~=~0.268 GeV/m and b~=~0.00047 m$^{-1}$ \cite{mmc}.  Cherenkov 
photons are propagated from the track assuming an effective scattering length of 21 m and an absorption length of 120 m.  We simulate a large number of photons and build photon density
tables as a function of distance from the track, valid for minimum ionizing muons and shown in \cite{method}.  We scale the photon normalization by the ratio of energy loss
to minimum energy loss (dE/dx/[0.268 GeV/m]) to obtain
the photon density for muon tracks with arbitrary energy loss.  We sample this density by Monte Carlo to determine PMT hits, assuming 10 inch diameter photomultiplier tubes with
20\% quantum efficiency.  Events with more than 14 modules hit satisfy our trigger and are kept.

Using the atmospheric neutrino fluxes of Barr {\it et al.} \cite{bartol}, we find such a detector will record 134,000 atmospheric neutrinos per year livetime.
We expect that about 50\% of the atmospheric
neutrinos will be retained after cuts necessary to reduce the background of downgoing muons from cosmic ray air showers, so we choose a sample of 67,000 events as the background to
compare our search methods.  We simulate transient point sources with E$^{-2}$ energy spectra using the same Monte Carlo.  The event angular offsets from the true muon
direction are sampled from a 2-D Gaussian with a median of 0.7$^{\circ}$, resulting in a median point spread of 0.86$^{\circ}$ when including the muon track deviation at the neutrino
interaction vertex.  Errors in muon energy estimation are assumed to be $\sim$0.3 in log$_{10}$E above $\sim$TeV \cite{amandareco, dima}, so we assign to each event a
reconstructed energy differing from the true energy by a value sampled from a Gaussian of width $\sim$0.3 in log$_{10}$E.  Below $\sim$1~TeV,
atmospheric neutrino fluxes are large with respect to the source signal, and hence energy information provides little signal separation power in this energy range.
Thus, neglecting the effect of worsening energy resolution below 1 TeV has negligible impact.  Signal event times are sampled from a Gaussian distribution,
with Gaussian widths ranging from 100 days to less then 10 ms.  

\subsection{A Neutrino Source with Known Time Dependence}

\begin{figure}\begin{center}
\mbox{\includegraphics[width=3.715in]{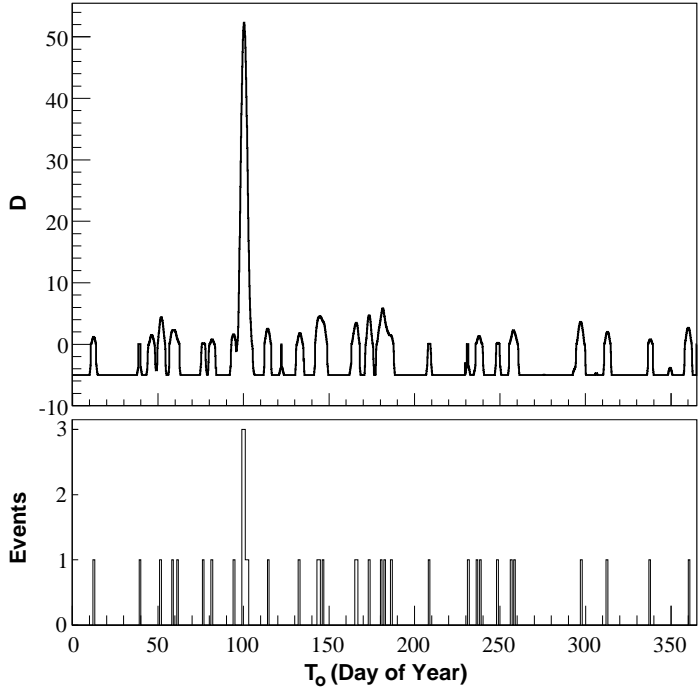}}
\caption{\label{Fig:TS} A simulated burst of 7 neutrino events with true maximum at day 100 and $\sigma_{\text{T}}$=1 day.  Top: Graph of $D$ as a function of $\text{T}_{\circ}$
over the simulated year of data, assuming we know $\sigma_{\text{T}}$=1 day, with peak very close to the true maximum at day 100.
Bottom: Events within 1.5$^{\circ}$ of the simulated source location, with signal events peaked at day 100 and background events distributed evenly throughout the period.}
\vskip 1cm
\end{center}\end{figure}
\begin{figure}\begin{center}
\mbox{\includegraphics[width=2.715in]{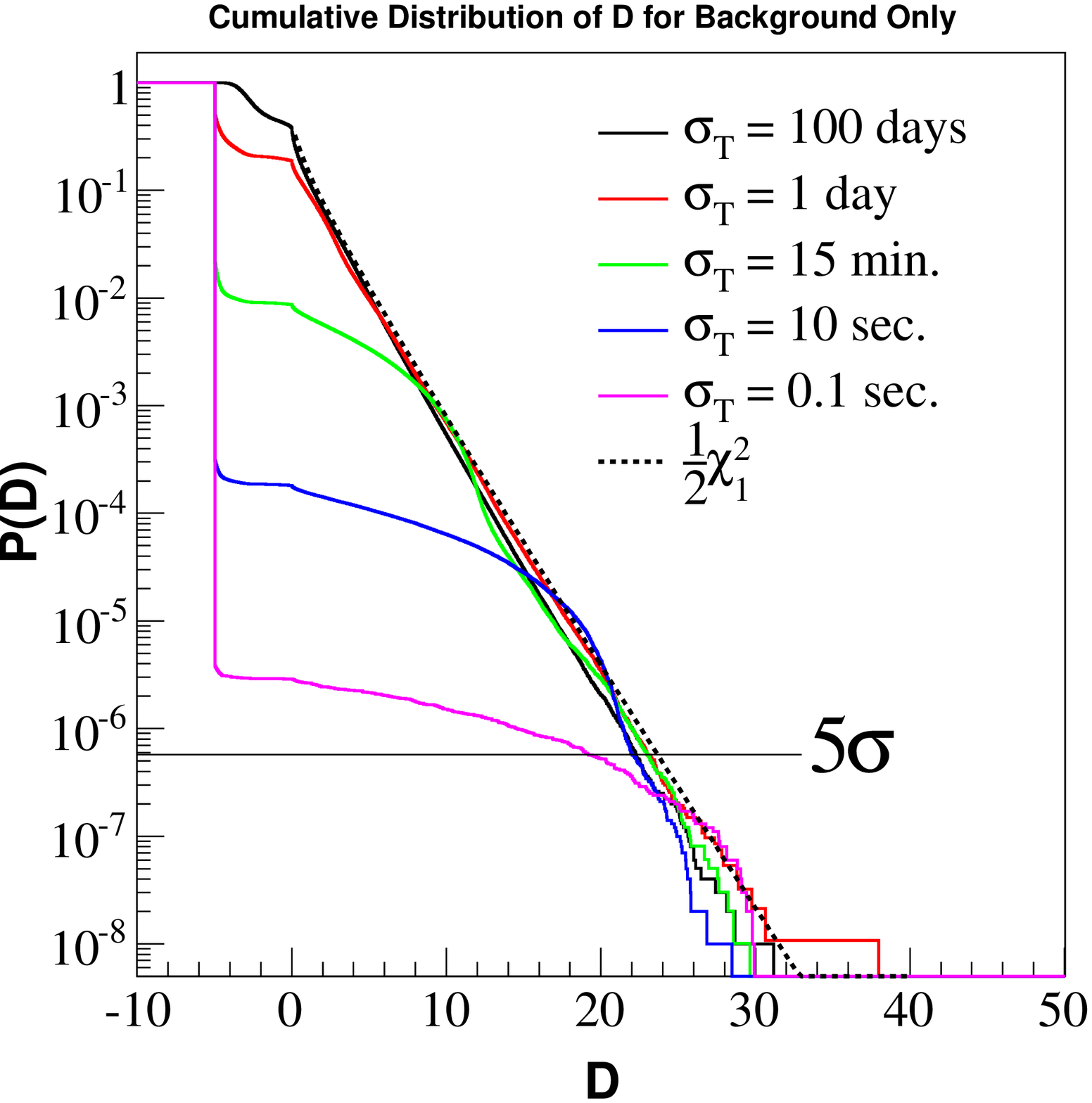}}
\mbox{\includegraphics[width=2.715in]{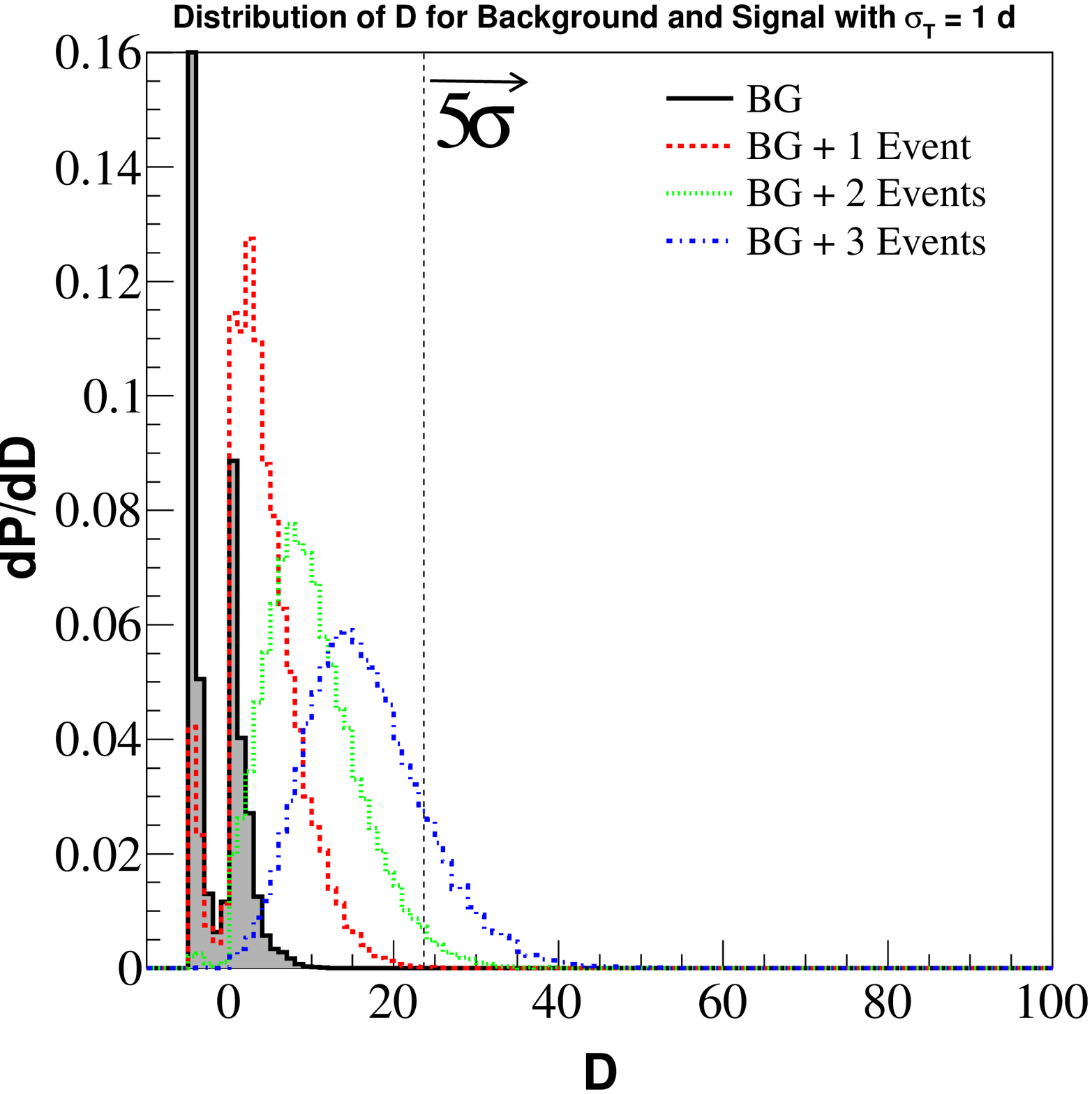}}\\
\mbox{\includegraphics[width=2.715in]{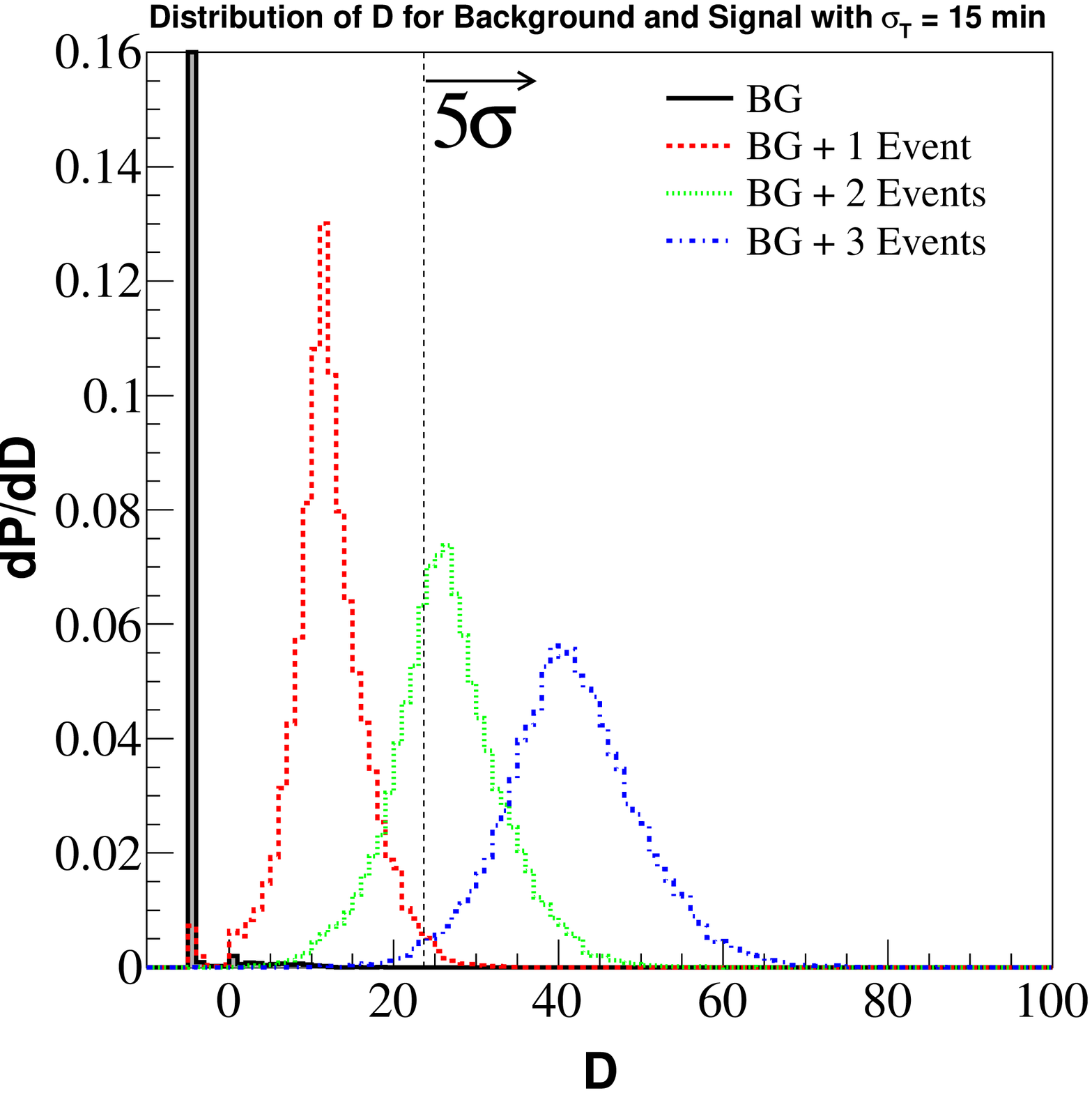}}
\mbox{\includegraphics[width=2.715in]{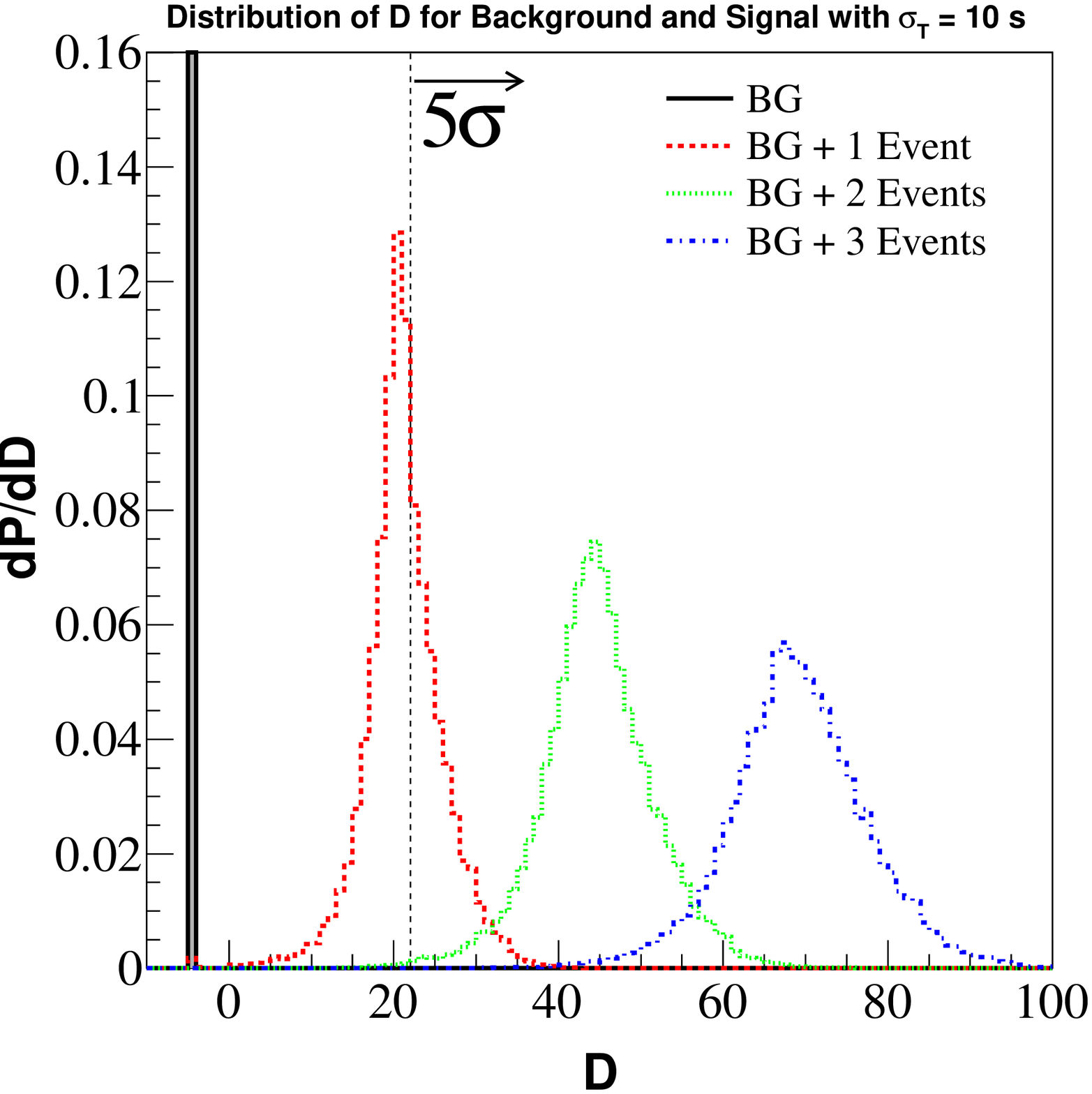}}
\caption{\label{Fig:FixedDist} Results for transient searches with known (fixed) T$_{\circ}$ and $\sigma_\text{T}$.  Top left: Background-only cumulative distributions of the
test statistic $D$ for Gaussian transients with durations $\sigma_\text{T}$ from 0.1 s to 100 days (solid lines from bottom to top respectively). For comparison, the $\chi^2$
distribution for 1 degree of freedom is shown with a dashed line.  Top right and bottom: Distributions of $D$ for background alone and
background with simulated Gaussian transients producing one, two, or three signal events and having width $\sigma_\text{T}$=1 day (top right), $\sigma_\text{T}$=15 minutes
(bottom left), and $\sigma_\text{T}$=10 seconds (bottom right).  In the upper right plot, the background distribution is double-peaked because trials with no nearby events
($D<0$) are pulled toward the fixed minimum value $D=-5$.}
\end{center}\end{figure}
The maximization of the likelihood in Eq. \ref{eq:lh} is performed numerically with MINUIT \cite{minuit}, using the MINIMIZE algorithm to minimize
$-2 \log \mathcal{L}(n_s, \gamma)$, yielding the test statistic $D$ of Eq. \ref{eq:knts}.  If no events are near the source
at the time of the burst, the test statistic $D$ tends to negative infinity, so we fix the lower bound of $D$ to $-5$.
In the top left plot of Fig. \ref{Fig:FixedDist}, the cumulative distribution of $D$ is shown for background alone, using a range of different fixed search widths $\sigma_{\text{T}}$. 
The distribution of $D$ is comparable to a chi square distribution with one degree of freedom for longer search windows when there are usually one or more events
correlated with the search window.  For shorter windows, on the other hand, the majority of background trials have no nearby correlated events, and therefore more frequently
result in the fixed minimum value of $-5$.
The remaining plots in Fig. \ref{Fig:FixedDist} show the test statistic distribution when one, two, or three signal events are added assuming Gaussian bursts of width
$\sigma_\text{T}$=1 day, $\sigma_\text{T}$=15 minutes, and $\sigma_\text{T}=10$ seconds in the top right, bottom left, and bottom right plots, respectively.
As signal events are added, \clearpage the distribution of $D$ shifts to larger values and separates from the background distribution.  Larger values of $D$ are
obtained for shorter values of $\sigma_\text{T}$ given a fixed number of signal events, demonstrating fewer events are needed to reach 5$\sigma$ for shorter burst durations.

\subsection{A Neutrino Source with Unknown Time Dependence}
\label{sec:resut}

We maximize the likelihood in equation \ref{eq:lh} with respect to $n_s$ and $\gamma$, and additionally
with respect to T$_{\circ}$ and $\sigma_\text{T}$ to identify a best-fit burst, yielding the test statistic $D$ of Eq. \ref{eq:uknts}.
Because of the large number of local maxima, shown in Fig. \ref{Fig:TS}, the numerical maximization of the likelihod by MINUIT cannot reliably find the global
likelihood maximum without accurate first guess values for the time parameters T$_{\circ}$ and $\sigma_\text{T}$.
To obtain first guess values, we first identify the set of events within 5$^{\circ}$ of the source location.  Then for successive values of $m$, starting with 
$m=2$, we treat each possible series of $m$ consecutive events in time as representing a Gaussian burst, with first guesses for T$_{\circ}$ and $\sigma_\text{T}$ given by
the mean time and RMS (relative to the mean time) of the events.  Fixing $n_s=m$ and $\gamma=2.0$, we maximize the likelihood, finding a temporary best-fit
$\hat{\text{T}}_{\circ}$ and $\hat{\sigma}_{\text{T}}$.  This is repeated for all possible consecutive series of $m$ events, and for all values of 2 $\leq m \leq$ 5.
For our simulation, we find that bursts with greater than 5 events are accurately identified with $m=5$, so we stop the search at this point.
For whichever combination that gives the overall highest likelihood, the corresponding best fit $\hat{\text{T}}_{\circ}$ and $\hat{\sigma}_{\text{T}}$ are then used as the
first guess for the final maximization with all four parameters ($n_s$, $\gamma$, T$_{\circ}$, $\sigma_\text{T}$) free.
Numerical maximization with MINUIT yields the global maximum likelihood and best fit parameters $\hat{n}_s$, $\hat{\gamma}$, $\hat{\text{T}}_{\circ}$, and $\hat{\sigma}_\text{T}$.

\begin{figure}\begin{center}
\mbox{\includegraphics[width=2.715in]{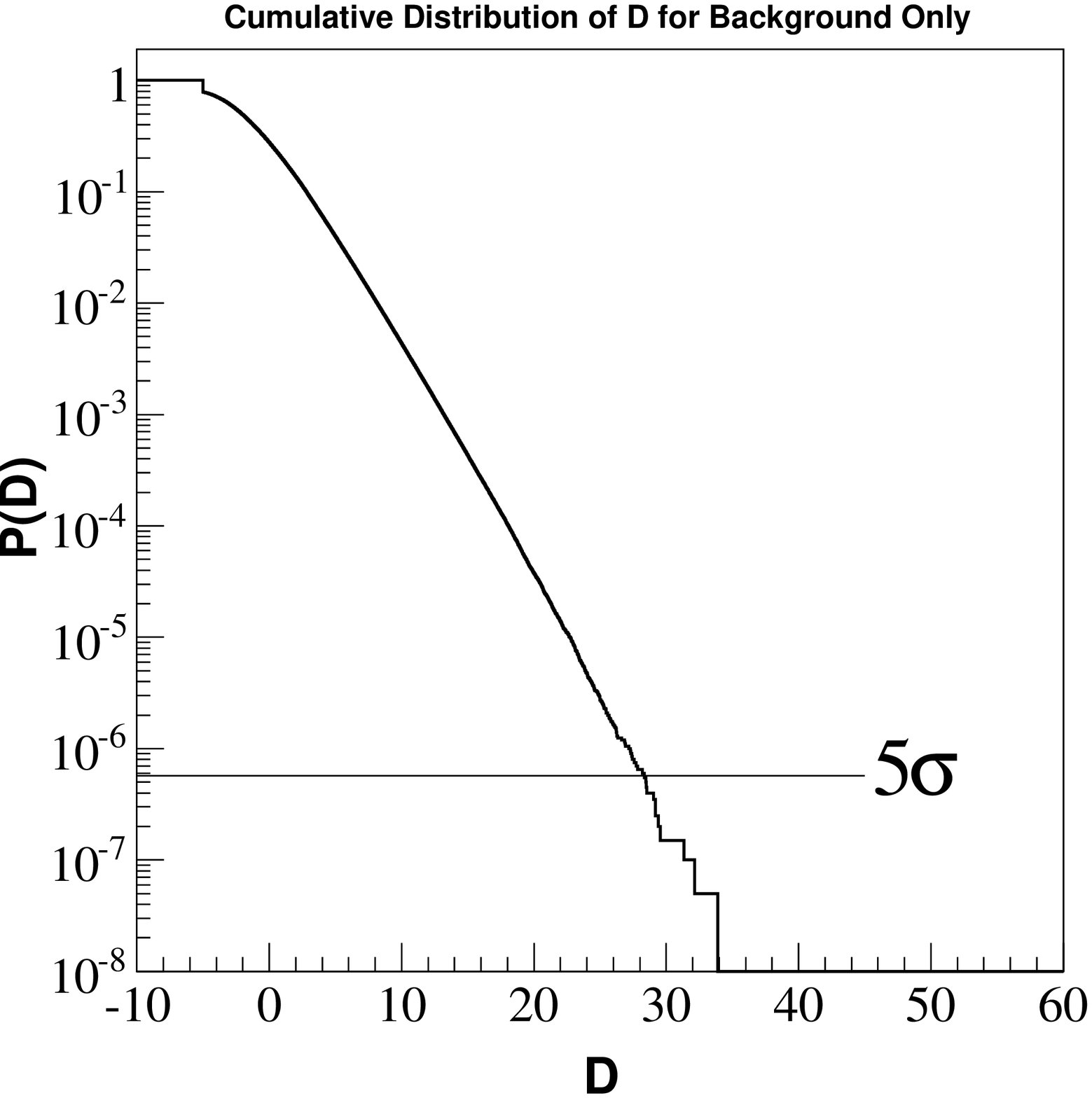}}
\mbox{\includegraphics[width=2.715in]{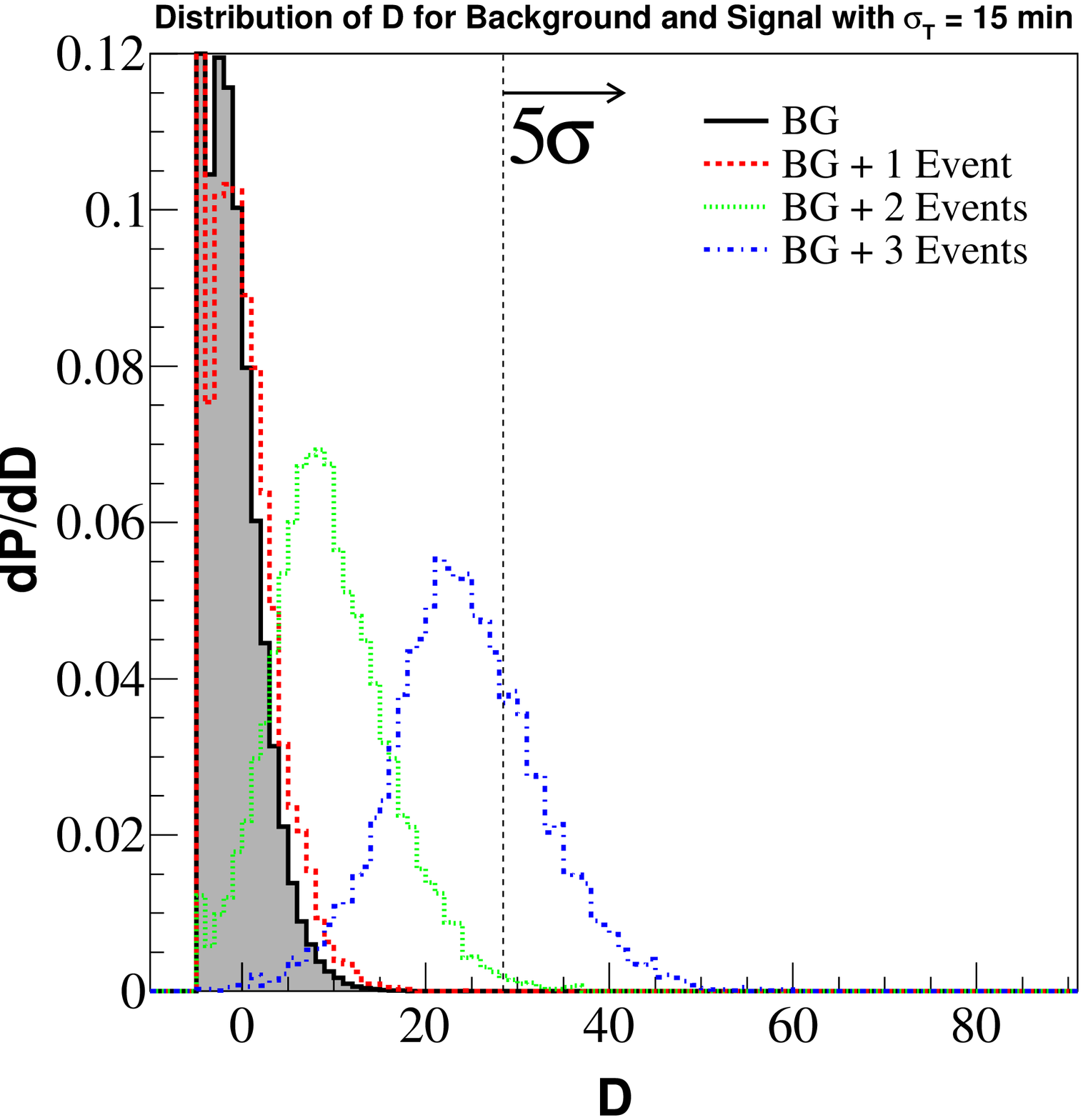}}\\
\mbox{\includegraphics[width=2.715in]{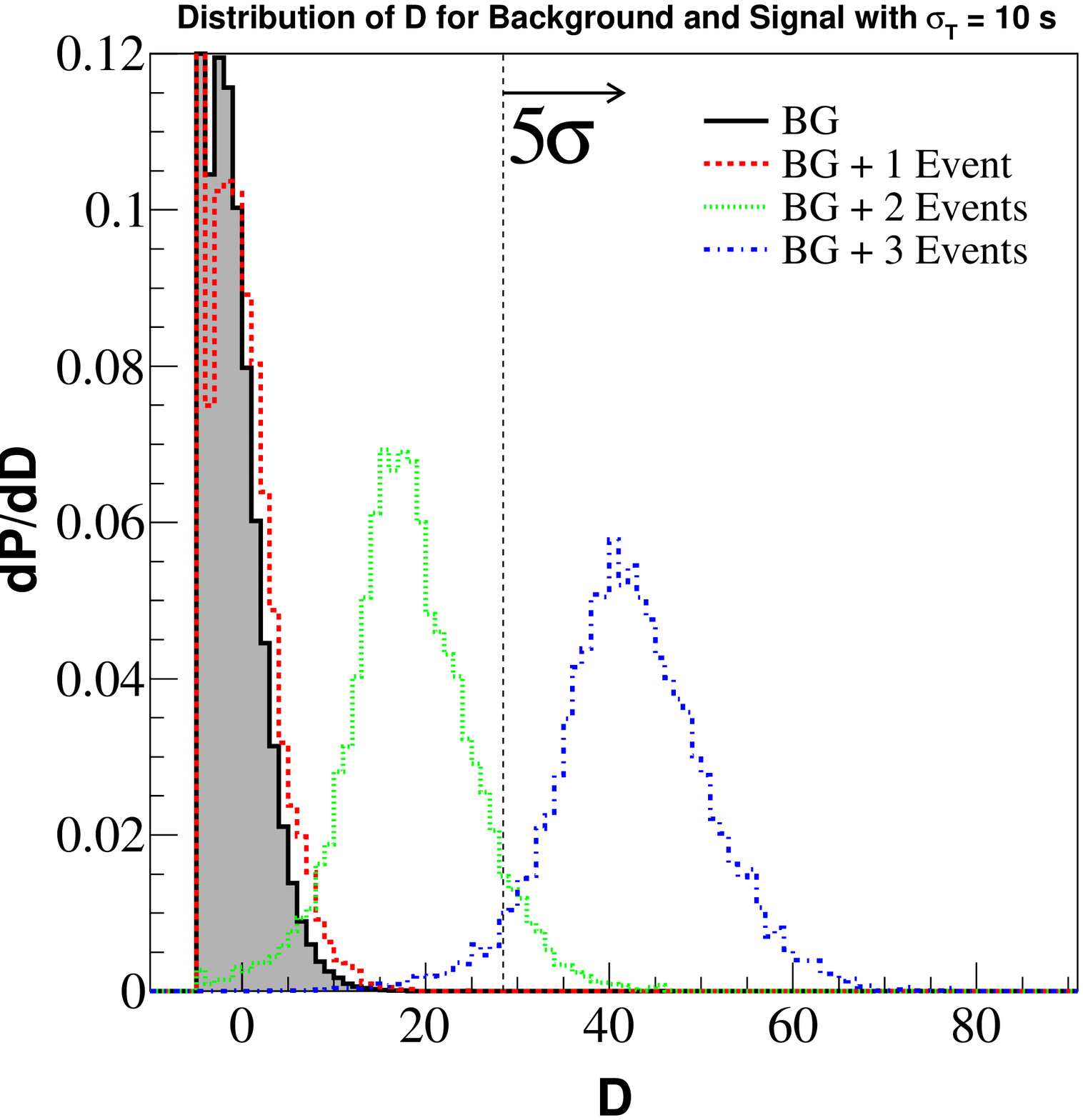}}
\mbox{\includegraphics[width=2.715in]{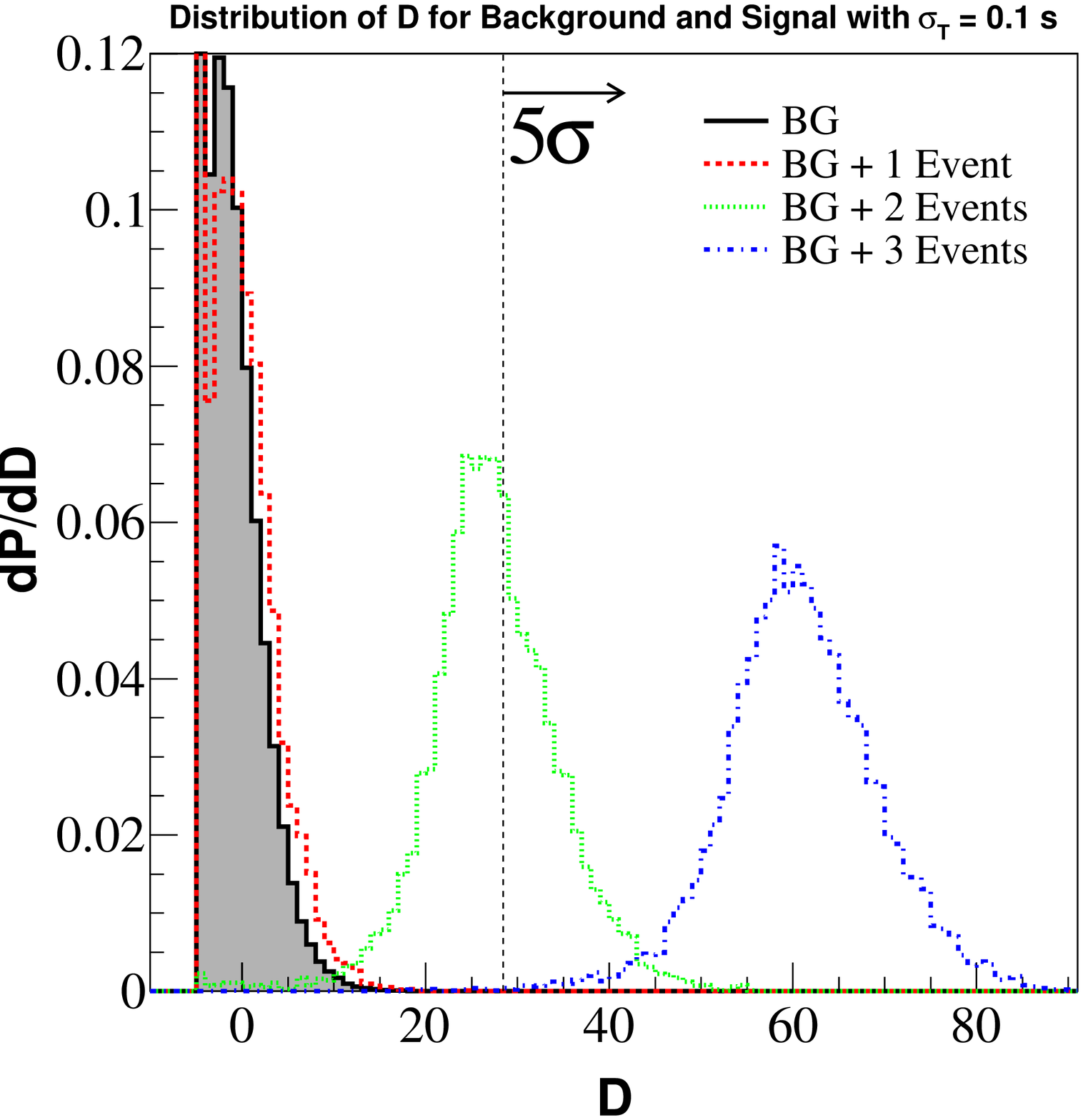}}\\
\caption{\label{Fig:varDist} Distributions of the test statistic when searching for Gaussian transients with an unknown time and duration.  Top left: Cumulative distribution of the
test statistic $D$ for background alone, indicating the $5\sigma$ level that corresponds to $D = 28.5$. Top right and bottom: Distributions of $D$ for background alone and
background with simulated Gaussian transients producing one, two, or three signal events and having width $\sigma_\text{T}$=15 minutes (top right), $\sigma_\text{T}$=10 seconds
(bottom left), and $\sigma_\text{T}$=0.1 seconds (bottom right).}
\end{center}\end{figure}

The distribution of the test statistic for randomized background alone is shown in the top left plot of Fig. \ref{Fig:varDist}.  This distribution does not follow a chi square.
For the present simulation, we find the 5$\sigma$ test statistic threshold to be $D$ = 28.5 after simulating a suitably large background sample.
A burst-like sequence of events is usually identified for some values of T$_{\circ}$ and $\sigma_\text{T}$, and distributions of $D$ for background trials are
more often centered near zero, rather than the minimum value of $-5$.  The remaining plots of Fig. \ref{Fig:varDist} show the distribution of $D$ for background
and one, two, or three added signal events from Gaussian transients of several durations, where the true burst time and duration are not known by the algorithm.
The distribution for background alone is not specific to any particular simulated width, and therefore is the same in each plot.  Similar to Fig. \ref{Fig:FixedDist},
larger values of $D$ are obtained for shorter values of $\sigma_\text{T}$ given a fixed number of 2 or 3 signal events.  A single signal event cannot be significant
by itself in an untriggered search, since generally there are several such events near the source over the range of time in the data set.  This is shown by the distributions
of $D$ with one added signal event in Fig. \ref{Fig:varDist}.

\section{Results}
\label{sec:results2}

\begin{figure}\begin{center}
\mbox{\includegraphics[width=11.6cm]{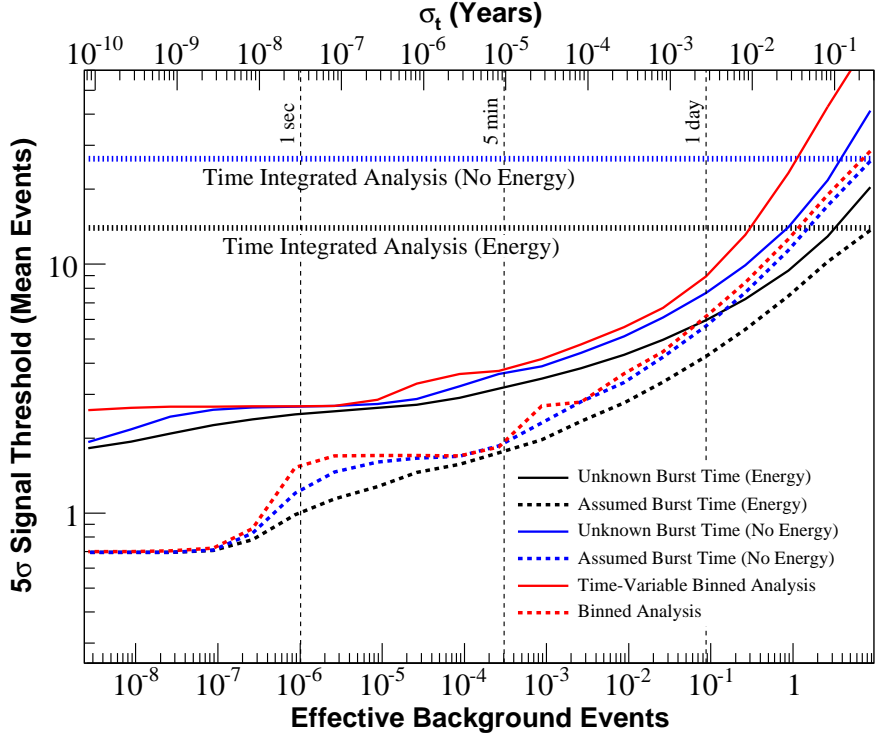}}
\caption{\label{Fig:DiscPot} Discovery potential of several search methods as a function of the number of background events within a bin of 1.35$^{\circ}$ radius and within
$\sigma_{\text{T}}$ of the burst time, which correlates to the upper axis of $\sigma_{\text{T}}$ for our simulated neutrino detector.
Shown are this method for searches with known (black dotted line)
and unknown (black solid line) time dependence, and this method without the event energy term for known (blue dotted line) and unknown (blue solid line) time dependence.  Also
shown are a binned search for bursts with known time dependence (red dotted line) and an untriggered time-variable binned search \cite{konstancja} (red solid line).}
\vskip 0.5cm
\end{center}\end{figure}

We apply the search to a range of burst durations, with Gaussian widths from 100 days to less than 10 milliseconds.  For each duration, we simulate 10,000
trials for each of 0--120 signal events added to the data set.  We also perform 10$^{8}$ trials using randomized background alone for each method, as shown in Figs.
\ref{Fig:FixedDist} and \ref{Fig:varDist}, to calculate the discovery potential at 5$\sigma$ for 50\% of trials.
This discovery potential is shown in Fig. \ref{Fig:DiscPot} as a function of the expected number of background events
within a bin of 1.35$^{\circ}$ radius and within $\sigma_{\text{T}}$ of the burst time.
Also shown is the discovery potential of the method without the energy term, and just the spatial and time terms in the PDF.  We compare to the time-independent
search and additionally to time-dependent binned methods.  For the case of known burst time and duration, we optimize bin angle and time cuts for each burst duration.  For
the binned search with unknown time dependence, we apply the method described in \cite{konstancja}, using variable time and bin angle cuts to identify the
most significant sequence of events.
For both binned methods, we calculate discovery potential at 5$\sigma$ for 50\% of trials using the same range of burst durations and range of signal events.

For a burst with $\sigma_{\text{T}}$ = 1 second, 10$^{-6}$ background events are expected in a bin of 1.35$^{\circ}$ radius and within $\sigma_{\text{T}}$ of the burst time.
The method with known time dependence requires on average $\sim$1 event for a 5$\sigma$ detection, while the binned method requires $\sim$1.5 events.  More events are necessary
for discovery when the time dependence is unknown due to trial factors, and on average $\sim$2.4--2.7 events are required to detect the same 1 second duration burst.
At the shortest timescales, only one event is necessary to discover the source in 50\% of trials (corresponding to a Poisson mean of 0.7 events) when the burst
time in known.  As the duration increases, the background increases proportionally, and more signal events are necessary for discovery.  The transitions from needing one event
to needing two events (at $\sim$5$\times 10^{-7}$ background events) and from two events to three events (at $\sim$7$\times 10^{-4}$ background events) are particularly sharp
for the binned search, resulting in the stairstep apperance in Fig. \ref{Fig:DiscPot}.  The unbinned method, especially when including the energy term, is less affected;
events which are less compatible with background, particularly high energy events, may push the test statistic above the 5$\sigma$ threshold and smooth the transitions.

For long-duration bursts
($\sigma_{\text{T}} > 0.1$~year), the untriggered search does not perform as well as the time-independent search.  For shorter bursts with durations lasting a few percent of our
simulated one year livetime, we find that the untriggered method using the energy term requires a factor of 2 fewer events for discovery than the time-independent search.
At the shortest timescales ($\sigma_{\text{T}} \leq 1$~second) a factor of 5 fewer events are required relative to the time-independent search.

\section{Conclusions}

We have described search methods for point sources with time dependent fluxes using a maximum likelihood approach, including a search for transients with an assumed time dependence
and an untriggered search for bursts when the burst time and duration are not known.  In the context of a neutrino point source search, we have calculated the discovery potential
of these methods as a function of the burst duration, demonstrating that our search methods require fewer signal events for detection than more traditional time-dependent binned
searches.  The method is generally applicable to point source searches in particle astrophysics, including gamma ray astronomy.

\section{Acknowledgments}

We thank the IceCube collaboration for thoughtful discussions.  We are also thankful to Alexander Kappes, Gary Hill and Juanan Aguilar for comments and suggestions.
We acknowledge support from the National Science Foundation -- Office of Polar Programs.

\end{document}